\documentstyle[12pt]{article}
\begin{document}
\title{A New Approach to Construct the Operator on Lattice for the
Calculation of
Glueball Masses}

\author{{Da Qing Liu$^2$, Ji Min Wu$^{1,~2}$ and Ying Chen$^2$}\\
        {\small $^1$CCAST(World Lab. ), P. O. Pox 8730, Beijing 100080,
                China}\\
        {\small $^2$Institute of High Energy Physics, Chinese Academy
                of
 Sciences,}\\
        {\small P. O. Box 918-4, Beijing, 100039, P. R. China}
       }
\maketitle

\begin{center}
\begin{minipage}{5.5in}
\vskip 0.8in
{\bf Abstract}\\
We develop a new approach to construct the operator on lattice for the
calculation of glueball mass, which is based on the connection
between the continuum limit of the chosen operator and the quantum
number $J^{PC}$ of the state studied. The spin of the state studied
in this approach is then determined uniquely and directly
in numerical simulation.
Furthermore, the approach can be applied to calculate the mass of
glueball states (ground or excited states) with any spin $J$ including $J\geq 4$.
Under the quenched approximation, we present
a pre-calculation result for the mass of $0^{++}$ state and
$2^{++}$ state, which are $1754(85)(86)MeV$ and $2417(56)(117)MeV$,
respectively.

\end{minipage}
\end{center}
\vskip 1in
\indent

\newpage
\section {Introduction}
During the past two decades,
extensive Monte Carlo simulations were carried out to calculate the glueball
mass spectra on lattice. $^{\cite{s1,s2,s3,s12,s13}}$ Most of
these papers carried out two key steps: one is the choice of glueball operators
with certain quantum number $J^{PC}$ basing on the method introduced in Ref.
\cite{s1} and the other is the application of
variational principle. Meanwhile, with a great amount of the improvement, such as
fuzzying and smearing, etc., these approaches surely work well and many
results of
well-controlled errors are obtained. However, the simulations are still puzzled
by an ambiguity, i.e., how to identify the definite spin $J$ from the irreducible
representation $R$ of the cubic point group, according to which the
glueball operators transform, because the corresponding between
this irreducible representation $R$ of cubic point group and spin $J$ is
not one-to-one.

Meanwhile, basing on the representation theory of $O(4)$ group and the
hypercubic
group, Mandula et. al.$^{\cite{s14}}$ developed an elegant scheme for the
choice of glueball
operators. With the definition of lattice gauge
field, they utilized the lattice color electric and magnetic
field to construct operators with definite $J^{PC}$
through the decomposition of the composition of lattice color electric and magnetic
field into certain representation of hypercubic group. But, the
correspondence between irreducible representation of hypercubic  group
and spin $J$ is also not one-to-one. The 'leading spin' is then assumed
when $a \rightarrow 0$.
But, this assumption cannot determine the spin uniquely.
For example,
one don't know how to separate 'leading spin' $J=1^{-}$ from
'leading spin' $J=2^{+}$ in ${\bf 6^{(+)}}$ representation and one can not
get the content of the non-leading spin$^{\cite{s14}}$.

We would like to show a possible solution to these troubles in this paper.
Unlike above references,  we start our discussion with the
asymptotic expansion of the operator. By expanding the chosen
operator according to power of lattice spacing $a$, we require
that the leading term of the expansion of the chosen operator belongs to
the irreducible representation $J^{PC}$ of $SO(3)^{PC}$ group.
We assume that the leading term of the expansion will give
the main effect to the state studied when $a \rightarrow 0$,
and the contribution should be  only given  by the leading term
in the continuum.
Therefore, the spin of the corresponding state is uniquely
determined by the leading term of the expansion when the lattice
 tends to continuum.

We claim that there are two advantages in this approach:
(1) Appending to the application of
variational method one can also study the contribution of
the different current ( with the same $J^{PC}$) to the definite state;
(2) One can determine the spin of the
corresponding state unambiguously and directly. In this way, we can
distinguish  state with definite $J^{PC}$ from other $J^{PC}$ states.

Some observations are shown in the forthcoming section.
We introduce our method in section 3
and give an example of pre-calculation to verify our method in section
4. Section 5 is a short summary.

\section {Some Observations}
In the continuum, the glueball states with definite quantum number
$J^{PC}$ make up of the bases of certain irreducible representation
$J^{PC}$ of $SO(3)^{PC}$ group. But, on lattice, there is only its
finite point subgroup, $O^{PC}$, and its corresponding irreducible
representations $R^{PC}(R=A_1, ~A_2, ~E, ~T_1$ and $T_2)$. Then to
measure glueball mass in lattice QCD, there arises such problem:
how we get correct results by only utilizing $O^{PC}$ group. To solve
the problem, authors make the continuum limit assumption($\beta=\infty$)
since Berg and Billoire$^{\cite{s1}}$, i.e., denoting masses of the
states extracted from operators in the irreducible representation $R$
by $m(R)$ and masses of the states with certain spin $J$ by $m(J)$,
they assume:$\cite{s1,s2,s12,s13}$
\begin{flushright}
$
\begin{array}{lr}
m(0^{PC})=m(A_1^{PC}),           & ~~~~~~~~~~~~~~~~~~~~~~~~~~~~(1a)\\
m(1^{PC})=m(T_1^{PC}),           & (1b)\\
m(2^{PC})=m(E^{PC})=m(T_2^{PC}), & (1c)\\
m(3^{PC})=m(A_2^{PC}).           & (1d)
\end{array}
$
\end{flushright}
But as Ref. \cite{s3} shows, this assumption is not always right.
From simulation results, in Ref. \cite{s3}, for example,
$T_1^{++}$ channel was not interpreted as
$J^{PC}=1^{++}$ states but most likely as $J^{PC}=3^{++}$ state ( less
likely
$J=6,7,9,... $ interpretation cannot rule out ),
since it seems this channel degenerated with $A_2^{++}$ channel
in the continuum.
As to the $T_1^{+-}$ channel, it was interpreted as $J^{PC}=1^{+-}$ state
instead of $3^{+-}$. But, $A_2^{+-}, T_1^{+-*} $ and $T_2^{+-}$ channels
become
degenereted in the continuum and they were interpreted as $3^{+-}$
state.
Therefore, we hope to develop an approach to determine spin $J^{PC}$ of
 states studied directly and uniquely in lattice simulation.

Meanwhile, on a $D=2+1$ lattice, Johnson and
Teper$^{\cite{s15}}$ found that in $A_1^{++}$ channel there exist two states
with different masses. They interpret the higher
one as $4^{++}$ state and lower one as $0^{++}$ state. Therefore, they
also consider that one needs a systematic and
general procedure to construct operators of arbitrary spin as
$a\rightarrow 0$$^{\cite{s15}}$.

Now, we present a possible procedure to solve these problems here.
Let us begin the discussion with some observations.\\

{\bf A.}\\
An arbitrary state $| \psi >$ can be regarded as generated
by current $o$ acting on vacuum $|0>$:
\setcounter{equation}{1}
\begin{equation}
|\psi>=o|0>.
\end{equation}
Since $|0>$ is invariance under Poincare group and $SU_C(3)$ group,
the character of $|\psi>$ can be described by $o$. For simplicity,
we only consider currents with mass dimension 4 here, saying
$B^a_i(x)B^b_j(x)$, where $B_i^a$ is color magnetic field and
$i,~j=1,~2,~3$.

Since $|\psi>$ is color singlet, we also require $o$ is color singlet.
One can get 6 color singlet currents from above combinations:
$$2Tr(B_i B_j)=\sum\limits_{a=1}^8 B_i^a B_j^a,$$
where $B_i=\sum\limits_{a=1}^8 B_i^a{\lambda^a \over 2}$.

We also require $|\psi >$ and $o$ transform as certain representation
$J^{PC}$ under $SO(3)^{PC}$ group. Since $\vec{B}$ transforms as $1^{+-}$
under this group, The $P,~C$ of $Tr(B_i B_j)$ are $++$.
Using C-G coefficients, we can decompose these bases consisting of
$Tr(B_iB_j)$ into $J=0$ and $J=2$( Due to color singlet,
there is no bases with $J=1$).
\footnote{ The systemic decomposition was well discussed by Jaffe
{\it et al}$^{\cite{s20}}$ in the study of the qualitative features of the
glueball spectrum. They suggest to construct glueball operators for
certain $J^{PC}$ states with color magnetic and electric fields in
the continuum case. But, we should go further to study the
construction of operators on lattice as the rest of the paper
points out.
}
Then, in the subduced
representation $J\downarrow O$ of the rotation group $SO(3)$
restricted to subgroup $O$, we find that the base in $J=0$ is the
basis of representation $A_1$, and we can further reduce another five
bases in $J=2$ according to irreducible representations $E$ and $T_2$
of the cubic point group. So, we can categorize these bases as:
\begin{flushright}
$
\begin{array}{llr}
J=0:~ & a_{11}=Tr(B_1 B_1+B_2 B_2+B_3 B_3);                        & (3a)\\
J=2:  & e_1=Tr(B_1 B_1-B_2 B_2),~e_2=Tr(B_1 B_1+B_2 B_2-2B_3B_3);  & (3b)\\
~~    & t_{21}=Tr(B_2 B_3),~t_{22}=Tr(B_1 B_3),~t_{23}=Tr(B_1 B_2).& (3c)
\end{array}
$
\end{flushright}
Here $a_{11}$ is the
basis of representation $A_1$, $e_1$ and $e_2$ construct bases of
representation $E$, while $t_{21},~t_{22}$ and $t_{23}$ just make up of
bases
of representation $T_2$.

This is just what table 3 ( see Appendix) tells us: the subduced
representation $J=2$ of the rotation group can be decomposing into
representation $E$ and $T_2$ in the cubic group; the subduced
representation $J=0$ is just representation $A_1$.

We can make similar analysis for higher mass-dimensional gauge invariant
operator consisting of color magnetic field and its covariant derivative.

{\bf B.}\\
Now, we consider how to construct a glueball operator on lattice.
By expanding the chosen operator according to
power of spacing $a$, we require that the leading term of the
expansion of the chosen operator belongs to and only belongs to
the irreducible representation $J^{PC}$ of $SO(3)^{PC}$ group.
This is the key point of this paper.
Therefore,
we first consider the perturbative expansion of Wilson loops
on lattice according to power of
lattice spacing $a$. A simple example is plaquette operator
(We denote unit vector in the positive $i$-direction by $\hat{i}$):
\setcounter{equation}{3}
\begin{equation} O_{ij}=\sum\limits_n
O_{ij}(n)=\sum\limits_n
Tr[1-U(n,i)U(n+\widehat{i},j)U^{-1}(n+\widehat{j},i)U^{-1}(n,j)],
\end{equation}
where the link variable $U$ is a connector and defined by
\begin{equation}
U(n,i)=P\exp(i \int^a_0 dt A_i(an+\hat{i}t),
\end{equation}
Where $P$ is path-order operator.

There are two methods
to expand the operator. One is the application of non-Abelian Stokes
theorem$^{\cite{s16,s17}}$ and the other one is introduced by Luscher and
Weisz in Ref.
{\cite{s4}}. We use both methods and get the same results:
\begin{eqnarray} \label{e1}
O_{ij}=&\sum\limits_n\{{a^4\over 2}Tr(F_{ij}F_{ij})(n)+{ia^6\over
6}Tr(F_{ij}F_{ij}F_{ij})(n)
\nonumber \\
          &-{a^6\over 24}Tr(F_{ij}(D_i^2+D_j^2)F_{ij})(n)\}+O(a^8),
\end{eqnarray}
where $F_{ij}=\partial_i A_j-\partial_j A_i-i[A_i,A_j]$ is gauge
field and $D_i \cdot=\partial_i-i[A_i,\cdot]$ is covariant
derivative.

We now consider the $PC=++$ sector of operators, or real part of
operators in Eq. (\ref{e1})
with ignoring the second term of r.h.s. in Eq. (\ref{e1}). Due to
$O_{ji}=O^*_{ij}$, there are three
non-zero independent operators $ReO_{12},~ReO_{23},~ReO_{31}$.
Restricting oneself into
the cubic group,  one can combine
these operators into representation $A_1^{++}$ and $E_2^{++}$:
\begin{eqnarray} \label{e2}
A_1^{++}:&
Re(O_{23}+O_{13}+O_{23})={a^4 \over 2}\sum\limits_nTr(B_1B_1+B_2B_2+B_3B_3)(n)+
O(a^6);
\nonumber \\
E^{++}:&
Re(O_{23}-O_{13})={a^4 \over 2}\sum\limits_nTr(B_1B_1-B_2B_2)(n)+O(a^6),
~~~~~~~~~
\nonumber \\
~&Re(O_{23}+O_{13}-2O_{12})={a^4\over 2}
 \sum\limits_nTr(B_1B_1+B_2B_2-2B_3B_3)(n)+O(a^6),
\end{eqnarray}
where color magnetic field is
$B_i=-{1\over 2}\sum\limits_{jk}\epsilon_{ijk}F_{jk}$.

We suppose that the leading term gives the most
contribution of the operator when $a$ is small enough, or, only the
leading term gives
the contribution in the continuum. While comparing Eq. (\ref{e2}) to Eq.
(3), it is assured that, in the continuum limit,
the state extracted from such operator $A_1^{++}$ is $J=0$, and the states
extracted from the operator in $E^{++}$ corresponds to $J=2$ state.

We should emphasis again,
in the general case, the continuum limits of
the operator in representation $E$ or in $T_2$ is not always corresponding
to $J=2$, i.e., the parallelism in Eq. (1) does not always hold. Only after
expanding the chosen operator as we do above, we
are then able to affirm or disaffirm the parallelism.

By the way, we should point out here that the non-leading terms
in the expansion of the operator do not always belong to the same $J^{PC}$
as that of leading term,
which will bring
up the mixing with different spin $J$. But, as argued above, we
expect that this artificial mixing will decrease with the
decreasing of lattice
spacing $a$ so that the mixing should vanish when $a\rightarrow 0$
although it will
affect our error estimate.
On the other hand, we can utilize non-leading terms to explore high-spin
states.

These two examples tell us that to calculate the mass of the definite
$J^{PC}$ state, we should require the continuum limits of our operators belong
to and only belong to $J^{PC}$ representation of $SO(3)^{PC}$ group. One can
get this aim by using the combination of the different operators which belong
to the same $R^{PC}$. We will present an example to construct the operator in
the following section and then show the simulation results and discuss the
errors in section 4.

\section {The Construction of the Operator}

We exemplify here how to construct operator $0^{++}$ and
$2^{++}$ up to $a^4$. On lattice, gauge-invariant current $o$ with
$0^{++}$ corresponding to
the scalar glueball can be written as
\begin{equation} \label{e7}
 o=\sum\limits_n
\{a^4\sum\limits_{i=1}^3 Tr(B_iB_i)(n) +a^6\times {\mbox (current~with
~mass~dimension~6)}+\cdots \},
\end{equation}
where the current with mass dimension 6 is the combination of
$\sum\limits_{i,j}^3Tr(D_iF_{ij}D_iF_{ij})$, $\sum\limits_{i,j,k}^3
Tr(D_iF_{jk}D_iF_{jk})$ and $\sum\limits_{i,j,k}^3
Tr(D_iF_{ik}D_jF_{jk})$. For simplicity, we only consider the
current $o$ up to mass dimension 4 in this paper. Then,
let us observe the sum of the planar special $2\times 1$ rectangular
over all lattice sites:
\begin{eqnarray}
O'_{ij} &=& \sum\limits_n O'_{ij}(n)
\nonumber \\
&=&{1\over 2}
\sum\limits_n  Tr\{[1-U(n,i)U(n+\hat{i},i) U(n+2\hat{i},j)
U^{-1}(n+\hat{i}+\hat{j},i)
\nonumber \\
& &U^{-1}(n+\hat{j},i)U^{-1}(n,j)]
+[1-U(n,i)U(n+\hat{i},j)
U(n+\hat{i}+\hat{j},j)
\nonumber \\
& & U^{-1}(n+2\hat{j},i)U^{-1}(n+\hat{j},j)U^{-1}(n,j)]\}.
\end{eqnarray}
The real part of the expansion for the operator $O'_{ij}$ is
\begin{eqnarray}
ReO'_{ij}&=&\sum\limits_n\{{a^4\over 2} 4Tr(F_{ij}F_{ij})(n)
\nonumber \\
 &&-{a^6\over24}10Tr(F_{ij}(D_i^2+D_j^2)F_{ij})(n)\}+O(a^8).
\end{eqnarray}
Then, we define
\begin{equation}
\Theta_{ij}(n)=Re(O_{ij}(n)-{1\over 10}O'_{ij}(n)).
\end{equation}
Continuum limit of operator $\Theta_{ij}$ is
\begin{equation}
\Theta_{ij}=\sum\limits_n{3a^4\over 10}Tr(F_{ij}F_{ij})+O(a^8).
\end{equation}
Decomposing $\Theta_{ij}$ into $A_1^{++}$ according to traditional method,
we get the basis of representation $A_1^{++}$:
\begin{eqnarray} \label{scalar}
F&\equiv &\Theta_{12}+\Theta_{13}+\Theta_{23}
\nonumber \\
&=&\sum\limits_n {3a^4\over 10}Tr(B_1B_1+B_2B_2+B_3B_3)+O(a^8).
\end{eqnarray}

Apparently, the quantum number of continuum limit of $F$ is $0^{++}$. In
other words, $F$ transforms as $0^{++}$ under $SO(3)^{PC}$ group up to
$a^4$. We
expect that the symmetry of $SO(3)$ has
been restored when $a\rightarrow 0$, the extracted state should be mainly
given
by the leading term of $F$,
so the extracted state is $0^{++}$ one in the continuum limit.
Operator $F$ is our aimed operator for $0^{++}$ state.

We may also choose the bases $G_1$ and $G_2$ of representation $E^{++}$ to
measure tensor glueball mass as follows. The operators and their expansion
are:
\begin{equation}
G_1=Re(\Theta_{23}-\Theta_{13})
   =\sum\limits_n{3a^4\over 10}Tr(B_1B_1-B_2B_2)+O(a^8),
\end{equation}
and
\begin{equation}
G_2=Re(\Theta_{23}+\Theta_{13}-2\Theta_{12})
   =\sum\limits_n {3a^4\over 10}Tr(B_1B_1+B_2B_2-2B_3B_3)+O(a^8).
\end{equation}
According to (3b), they belong to bases of representation $2^{++}$ up to
$O(a^4)$.

\section {Simulation Results}
Under the quenched approximation, we perform our calculation on
anisotropic
lattice with improved gluonic action as chosen in Ref. $^{\cite{s5}}$
\begin{equation}
S_{II}=\beta \{ {5\Omega_{sp}\over 3\xi u_s^4}+{4\xi\Omega_{tp}\over 3u_s^2u_t^2}
-{\Omega_{sr}\over 12 \xi u_s^6}-{\xi\Omega_{str}\over 12u_s^4u_t^2} \},
\end{equation}
where $\beta=6/g^2,~g$ is the QCD couple constant, $u_s$ and $u_t$ are mean link
renormalization parameters(we set $u_t=1$), $\xi=a_s/a_t$ is the aspect ratio,
and  $\Omega_{sp}$ includes the sum over all spatial plaquettes on the
lattice, $\Omega_{tp}$ indicates the temporal plaquettes, $\Omega_{sr}$
denotes the planar $2\times 1$ spatial rectangular loops and $\Omega_{str}$
refers to the short temporal rectangles( one temporal and two spatial links).
More detail is given in Ref. \cite{s5}. In each $\beta$ calculation, we set
2800 sweeps to make configurations reach to equilibrium and make
measurement once after
every four sweeps. Our calculation spends 80 bins in which there are 70 measurements after
reaching equilibrium. The method to set the scale used here is introduced by
ref \cite{s8}. For each $\beta$, we have measured $u_s^4$ and found they
coincide with those in Ref. \cite{s3} in the error bound. So we adopt the
data in Ref. \cite{s3} as our energy scale. Table 1
shows the simulation parameters\\
\begin{center}
\begin{tabular}{|c|c|c|c|c|c|} \hline
$\beta$ &$\xi$ &$u_s^4$ &Lattice         &$r_s/r_0$ &$a_s(fm)$  \\ \hline
1.7     & 5    &0.295   &$6^3\times 30$  &0.8169   & 0.39      \\ \hline
1.9     & 5    &0.328   &$8^3\times 40$  &0.727    & 0.35      \\ \hline
2.2     & 5    &0.378   &$8^3\times 40$  &0.5680   & 0.27      \\ \hline
2.4     & 5    &0.409   &$8^3\times 40$  &0.459    & 0.22      \\ \hline
2.5     & 5    &0.424   &$10^3\times 50$ &0.407    & 0.20      \\ \hline
\end{tabular} \\
{\small {\bf Table 1} The glueball simulation parameters$^{\cite{s3}}$. Here we
assume $r_0=410(20)MeV$.}
\end{center}

As argued above, we choose operator $F$ to calculate scalar
glueball mass and operator $G_1$ and $G_2$ to calculate tensor
glueball mass. As usual, we calculate the
average over sample vacua of a correlation $C(t)=<0|o^R(t)o^R(0)|0>$, where
$o^R(t)=o(t)-<0|o(t)|0>$ is the vacuum-subtracted form of the
chosen operator, to determine masses of the
corresponding glueball states with the improvements such as
fuzzying and smearing. At the same time, following the mean
field theory$^{\cite{s6}}$, we also replace link variant $U$
by $U/u_s$ in the chosen operators due to the tadpole
correction. After such programmes, we get our results
in each $\beta$ which are shown in table 2:
\begin{center}
\begin{tabular}{|c|c|c|c|c|c|} \hline
 $ \beta$ & 1.7   &1.9   &  2.2 &  2.4     &    2.5  \\ \hline
scalar    &0.609(4) &0.515(8) &0.412(7)  &0.315(6) & 0.322(2)) \\ \hline
tensor    &1.019(3)   &0.95(1)  &0.71(2) &0.548(6)  &0.519(4)  \\ \hline
\end{tabular} \\
{\small {\bf Table. 2}~ Glueball energy $ m_{G} a_t$ for each $\beta$.\\
The numerals in the brackets are the error estimates.
 }
\end{center}

Now we comment a little on the error estimate.

First, our action breaks the rotation symmetry to $O(a_s^4,a_t^2)$, i.e., the
upper limit of the precision in the calculation is $O(a_s^4,a_t^2)$. Since
as argued by many authors, the contribution of $O(a_t^2)$ can be ignored, the
upper limit of the precision here is $o(a_s^4)$.

Second, we ignore terms(currents) with mass dimension 6 in Eq. (\ref{e7}). Due to
dimensional analysis, the contribution of the terms to error should have a
square mass suppression$^{\cite{s20}}$, which will make two effects on our mass
measurement. One is that we should include it in systematic error in the
continuum limit, which needs further calculation to get its accurate value.
Here we simply expect that it is about $({\Lambda_{QCD}/M})^2$,
where we set
$\Lambda_{QCD}\simeq 250 MeV$ and $M$ is measured mass. The
second is that it will takes $O(a_s^2)$ error when
$a_s\neq 0$.
Since it is
not statistical error, its contribution to error can be fitted by
$c_2a_s^2+c_4a_s^4+\cdots$.

From the argument and calculated data, we use the formula
$m(0^{++},a_s)=1.754 -1.514(a_s/r_0) + 1.773 (a_s/r_0)^2$
 and
$m(2^{++},a_s)=2.417+0.783(a_s/r_0)^2-0.787(a_s/r_0)^4$
( unit: $GeV$) to fit our data.
We present our data and fitting curves in Fig. 1.

The statistical error is $0.076GeV$ for scalar glueball and $0.044GeV$
for tensor
glueball. According to Ref. \cite{s3}, systematic error is 1 percent
 ( from aspect ratio ). But since
our method also gives about 2 and 1 percent systematic error for $0^{++}$
and $2^{++}$ states respectively, the total
systematic error is about 2.2 percent( $39MeV$ ) and 1.4 percent( $34MeV$ )
respectively.
Therefore, the mass of scalar glueball is $1.754(85)GeV$ and the mass of
tensor glueball is $2.417(56)GeV$. Including the uncertainty in
$r_0^{-1}=410(20)Mev$, Our final results are:
$M_G(0^{++})=1754(85)(86)MeV$ and $m_G(2^{++})=2417(56)(117)MeV$.

\section{Conclusion}
Basing on the connection of the asymptotic expansion of the operator
and the quantum number $J^{PC}$ of the extracted state,
we present a new approach to construct operator on lattice for the
calculation of the glueball mass, which may solve the ambiguity in
the simulation. This approach points out that, in general, to
calculate the  mass of definite $J^{PC}$ glueball states,
first one should write out these currents
which transform as the representation $J^{PC}$ under
the $SO(3)^{PC}$ group in continuum and decompose them into irreducible
representations
$R^{PC}$ of the group $O^{PC}$ in the
subduced representation which is obtained by trivially embedding the
$O^{PC}$ group into the $SO(3)^{PC}$ group; then one should construct
corresponding operators which belongs to the representation $R^{PC}$ of
the $O^{PC}$ group
on lattice, its continuum limits
should be those currents mentioned above.
% next one should also use the variantional procedure to extract masses
% of the ground $J^{PC}$ states and other excited $J^{PC}$ states.

To verify our approach, we
calculate the scalar and tensor glueball mass under the quenched
approximation in this approach. Since the continuum limit of operator
$F$ is
actual $0^{++}$, we affirm the mass extracted from the operator
$F$ is scalar glueball mass and its value is $1754(85)(86)MeV$.
With the same reason, the mass extracted from operator $G_i(i=1,2)$ is
the tensor glueball one and its value is $2417(56)(117)MeV$.
These results are consistent with those obtained in references
${\cite{s3,s13,s5,s21}}$.

Apparently, there is no radical obstacle to prevent us to calculate
the mass of states with any spin $J $ including $J\geq 4$ in this
approach. We will make such study systematically. We have
first calculated the
mass of the ground $4^{++}$ states in the $E^{++}$ channel
with $2^{++}$ stete under the quenched
approximation in this approach.
It will be shown elsewhere.

Of course, the operator, its continuum limit transform as $J^{PC}$
of $SO(3)^{PC}$ group, is not unique. For example, one can also construct
the operator including color electric field. With these
operators, one can determine their relative weights of contribution by
variational principle. But, we did not make such treatment here.

To compare our results with the experiments, we need to calculate it
on an unquenched lattice. We should also study the mixing between
glueball states and normal mesons with the same $J^{PC}$. Such
works should be finished in future.

\section *{Acknowledgements}
We would like to thank Prof. Z. Q. Ma, T. Huang and K. F. Liu for the helpful
discussions and useful comments.

This works is supported by the National Natural Science Fundation of China
under the Grant No. 19991487, 10075051, and by the National Research
Center for Intelligent Computing System under the contract No.
99128, 00132.

\newpage

\section * {Appendix}
In the cubic group, the subduced representation $J$ of irreducible
representation $J$ can be decomposed into irreducible representation $R$.
Table 3 shows the multiplicity of decomposing $J$ up to $J=6$. It is well
known that the subduced representation with $J\geq 2$ are reducible and only
up to $J=3$ do new irreducible representation of the cubic group show up. \\
\begin{center}
\begin{tabular}{|c|c|c|c|c|c|c|c|}\hline
$R\backslash J$ &~0~ & ~1~ & ~2~ &~3~ &~4~ &~5~ &~6~ \\ \hline
$A_1$  & 1  & 0  & 0   & 0  &  1  & 0   &   1   \\ \hline
$A_2$  & 0  & 0  & 0   & 1  &  0  & 0   &   1   \\ \hline
$E$    & 0  & 0  & 1   & 0  &  1  & 1   &   1   \\ \hline
$T_1$  & 0  & 1  & 0   & 1  &  1  & 2   &   1   \\ \hline
$T_2$  & 0  & 0  & 1   & 1  &  1  & 1   &   2   \\ \hline
\end{tabular} \\

{\small {\bf Table 3}~~~ The Subduced representations of rotation group \\
up to $J=6$ for the cubic group.}\\
\end{center}

\vskip 1.0in
\section* {Figure Caption}
{\bf Figure 1}~~~~Masses of scalar and tensor glueball against the lattice spacing
$(a_s/r_0)^2$. The fitting curve are
$m(0^{++},a_s)=1.754-1.514(a_s/r_0)^2+1.773(a_s/r_0)^4$ for scalar
glueball mass and
$m(2^{++},a_s)=2.417+0.783(a_s/r_0)^2-0.787(a_s/r_0)^4 (unit:GeV)$
for tensor glueball mass. The masses in continuum limit are
$1.754(76)GeV$ and $2.417(44)GeV$ if
we only consider the statistical error. The top data and curve:
$m(4^{++},a_s)=3.65-1.22(a_s/r_0)^2+2.74(a_s/r_0)^4 (unit:GeV)$ is
for $4^{++}$ glueball mass.


\newpage
\begin{thebibliography}{99}
\bibitem{s1}
B. Berg and A. Billoire, Nucl. Phys. {\bf B}221(1983) 109.
\bibitem{s2}
UKQCD collabration et al, Phys. Lett. {\bf B}309(1993) 378.
\bibitem{s3}
C.J. Morningstar and M. Peardon, Phys. Rev. {\bf D}60(1999)
034509.
\bibitem{s12}
C. Micheal and M. Teper, Nucl. Phys. {\bf B}314(1989) 347.
\bibitem{s13}
H. Chen $et~al$, Nucl. Phys. {\bf B}(Suppl)34(1994) 357.
\bibitem{s4}
M. Luscher and P. Weisz, Commun. Math. Phys. 97(1985) 59.
\bibitem{s16}
O. Nachtmann: "High Energy Collision and Nonperturbative QCD" in
the lecture notes at the workshop "Topics in Field Theories" held
on Oct. 1993, Kloster Banz, Germany.
\bibitem{s17}
D. Q. Liu and J. M. Wu, hep-lat/0104007.
\bibitem{s5}
C. J. Morningstar and M. Peardon Phys. Rev. {\bf D}56(1997)
4043-4061.
\bibitem{s6}
G.P. Lepage and P. B. Mackenizie, Phys. Rev. {\bf D}48(1993)
2250-2264.
\bibitem{s8}
R. Sommer, Nucl. Phys. {\bf B}411(1994) 839-854.
\bibitem{s14}
J. E. Mandula $et~al$, Nucl. Phys. {\bf B}228(1983) 109.
\bibitem{s15}
R. Johnson and M. Teper, Nucl. Phys. {\bf B}( Suppl.) 73(1999) 267.
\bibitem{s20}
R. L. Jaffe, K. Johnson and Z. Ryzak, Ann. Phys., {\bf 168}(1986)
344.
\bibitem{s21}
C. Liu, Chin. Phys. Lett. {\bf 18}(2001) 187.
\end{thebibliography}
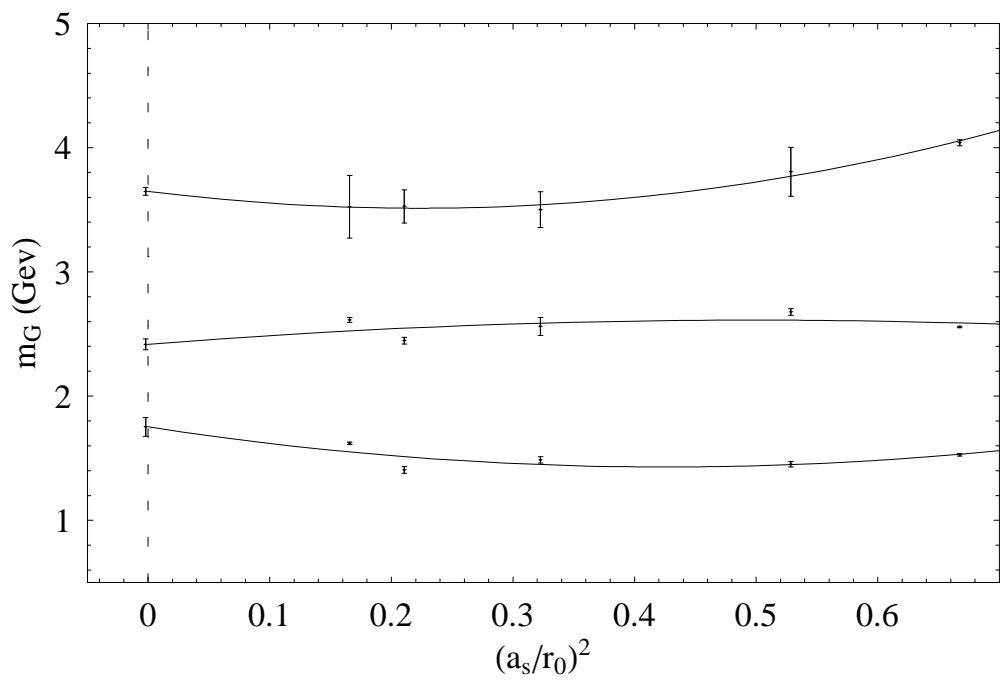
\end{document}